\documentclass[superscriptaddress, twocolumn, prd, mediabb]{revtex4}

\usepackage{amssymb,amsmath,array,verbatim}
\usepackage[dvipdfmx]{graphicx}

\newcommand{\beq}{\begin{eqnarray}}
\newcommand{\eeq}{\end{eqnarray}}
\newcommand{\p}{\partial}

\newcommand{\hs}[1]{\hspace{#1 mm}}
\newcommand{\bpm}{\begin{pmatrix}}
\newcommand{\epm}{\end{pmatrix}}
\newcommand{\Z}{\mathbb{Z}}

\newcommand{\C}{\mathbb{C}}

\newcommand{\D}{\mathcal D}

\newcommand{\ba}{\left(\begin{array}}
\newcommand{\ea}{\end{array} \right)}

\begin{document}

\title{
Exact Resurgent Trans-series and Multi-Bion Contributions to All Orders
}

\author{Toshiaki Fujimori}
\email{toshiaki.fujimori018(at)gmail.com}
\address{Department of Physics, 
and Research and 
Education Center for Natural Sciences, 
Keio University, 4-1-1 Hiyoshi, Yokohama, Kanagawa 223-8521, Japan
}

\author{Syo Kamata}
\email{skamata(at)rikkyo.ac.jp}
\address{Physics Department and Center for Particle and Field Theory, Fudan University,
220 Handan Rd., Yangpu District, Shanghai 200433, China
}

\author{\\ Tatsuhiro Misumi}
\email{misumi(at)phys.akita-u.ac.jp}
\address{Department of Mathematical Science, Akita 
University, 1-1 Tegata-Gakuen-machi,
Akita 010-8502, Japan
}
\address{Department of Physics, 
and Research and 
Education Center for Natural Sciences, 
Keio University, 4-1-1 Hiyoshi, Yokohama, Kanagawa 223-8521, Japan
}

\author{Muneto Nitta}
\email{nitta(at)phys-h.keio.ac.jp}
\address{Department of Physics, 
and Research and 
Education Center for Natural Sciences, 
Keio University, 4-1-1 Hiyoshi, Yokohama, Kanagawa 223-8521, Japan
}

\author{Norisuke Sakai}
\email{norisuke.sakai(at)gmail.com}
\address{Department of Physics, 
and Research and 
Education Center for Natural Sciences, 
Keio University, 4-1-1 Hiyoshi, Yokohama, Kanagawa 223-8521, Japan
}

\begin{abstract}
The full resurgent trans-series is found exactly in 
near-supersymmetric ${\mathbb C}P^{1}$ quantum mechanics. 
By expanding in powers of the SUSY breaking deformation 
parameter, 
we obtain the first and 
second expansion coefficients of the ground state energy. 
They are absolutely convergent series of nonperturbative 
exponentials corresponding to multi-bions with perturbation 
series on those backgrounds. 
We obtain all multi-bion exact solutions for finite time 
interval in the complexified theory. 
We sum the semi-classical multi-bion contributions that reproduce 
the exact result supporting the resurgence to all orders. 
We also discuss the similar resurgence structure in ${\mathbb C}P^{N-1}$ ($N>2$) models.
This is the first result in the quantum mechanical model where 
the resurgent trans-series structure is verified to all orders 
in nonperturbative multi-bion contributions.
\end{abstract}

\maketitle


\section{Introduction}
Path-integral has been extremely useful in many areas of 
quantum physics through perturbative and nonperturbative 
analysis. 
It is crucial to understand contributions from all the complex saddle points based on the thimble analysis in the path integral in order to give a proper foundation of quantum theories.
The resurgence theory gives a stringent relation between 
a divergent perturbation series and a nonperturbative exponential 
term, which often allows reconstruction from each other
\cite{Bogomolny:1980ur,ZinnJustin:1981dx,ZinnJustin:1982td,
ZinnJustin:1983nr, Voros1}. 
Resurgence is originally developed in studying ordinary 
differential equations and provides a trans-series, containing 
infinitely many nonperturbative exponentials and divergent 
perturbation series \cite{Ec1}. 
The intimate relation between these infinitely many nonperturbative 
contributions and perturbative ones is expected to provide an 
unambiguous definition of quantum theories. 
A mathematically rigorous foundation of path integral is 
now envisaged \cite{K1,K2,K3}. 
Resurgence has been most precisely studied recently in 
quantum mechanics (QM) to yield relations between nonperturbative 
and perturbative contributions systematically\cite{Basar:2013eka,
ZinnJustin:2004ib,ZinnJustin:2004cg,Jentschura:2010zza,
Jentschura:2011zza,Dunne:2013ada,Alvarez1,Alvarez2,Alvarez3, Dunne:2014bca,Misumi:2015dua,
Gahramanov:2015yxk,Dunne:2016qix,Behtash:2015loa,Behtash:2015zha,
Fujimori:2016ljw,Sulejmanpasic:2016fwr,Dunne:2016jsr,Kozcaz:2016wvy,Basar:2017hpr}, 
2D quantum field theories (QFT)\cite{Dunne:2012ae,Dunne:2012zk,
Cherman:2013yfa,Cherman:2014ofa,Misumi:2014jua,Misumi:2014bsa,
Misumi:2014rsa,Nitta:2014vpa,Nitta:2015tua,Behtash:2015kna,
Misumi:2016fno,Sulejmanpasic:2016llc}, 4D QFT\cite{Unsal:2007vu,Unsal:2007jx,
Shifman:2008ja,Poppitz:2009uq,Anber:2011de,Poppitz:2012sw,Argyres:2012vv}, 
supersymmetric (SUSY) gauge theories\cite{Aniceto:2014hoa,
Dunne:2015eoa,Aniceto:2015rua,Dorigoni:2015dha,Honda:2016mvg}, the matrix 
models and topological string theory \cite{Marino:2006hs,Marino:2008ya,Marino:2007te,Pasquetti:2009jg,Garoufalidis:2010ya,Aniceto:2011nu,Aniceto:2013fka,Santamaria:2013rua,Buividovich:2015oju}.

In the resurgent trans-series for theories with degenerate vacua, 
one needs to take account of configurations called ``bions" 
consisting of an instanton and an anti-instanton 
\cite{ZinnJustin:1981dx, ZinnJustin:2004ib}, which 
give imaginary ambiguities cancelling those of 
non-Borel-summable perturbation series. 
Recently single bion configurations are identified as 
saddle points in the complexified path integral 
\cite{Behtash:2015zha}. 
Exact solutions of the holomorphic equations of motion (complex 
and real bion solutions) are found in the complexified path 
integral of double-well, sine-Gordon and ${\mathbb C}P^{1}$ 
quantum mechanical models with fermionic degrees of freedom 
(incorporated as the parameter $\epsilon$) \cite{Behtash:2015zha, 
Fujimori:2016ljw}. 
${\mathbb C}P^{1}$ quantum mechanics is a dimensional 
reduction of the two-dimensional ${\mathbb C}P^{1}$ sigma model, which shows
asymptotic freedom, dimensional transmutation and the existence of instantons akin to four-dimensional QCD.
Contributions from these solutions are evaluated based on 
Lefschetz-thimble integrals and it is shown that the combined 
contributions vanish for the SUSY case $\epsilon = 1$, in 
conformity with the exact results of SUSY \cite{Fujimori:2016ljw}.
On the other hand, for the non-SUSY case $\epsilon \not= 1$,
the result contains the imaginary ambiguity, which is expected 
to be cancelled by that arising from the Borel resummation of 
perturbation series. 

Trans-series generically contain high powers of 
nonperturbative exponential, which may correspond 
to multiple bions. 
Non-SUSY models including ${\mathbb C}P^{N-1}$ quantum 
mechanics 
have been worked out explicitly to several low orders, but it 
was difficult to reveal explicitly the full trans-series to all 
powers of nonperturbative exponential and to ascertain their 
resurgence structure\cite{Dunne:2014bca,Misumi:2015dua,ZinnJustin:1981dx}. 
Localization in SUSY models helped to uncover the full trans-series, 
but so far their resurgence structures are found to be trivial 
without imaginary ambiguities\cite{Aniceto:2014hoa, Honda:2016mvg}.

The purpose of this work is to present and to verify the 
complete resurgence structure of the trans-series in 
${\mathbb C}P^{1}$ QM (and partly ${\mathbb C}P^{N-1}$ QM), 
focusing on the near-SUSY regime $\epsilon\approx 1$ where 
we can obtain exact results which exhibit resurgence structure 
to infinitely high powers of nonperturbative exponential. 
We will show that the contributions from an infinite tower of 
multi-bion solutions yield all these nonperturbative exponentials.
This is the first result revealing the thimble structure of all the complex saddle points, which is useful not only to understand the resurgence structure in quantum theories but also to study complex path integrals including real-time formalism and finite-density systems in condensed and nuclear matters \cite{Witten:2010cx, Cristoforetti:2013wha, Fujii:2013sra, Tanizaki:2014tua, Tanizaki:2014xba, Alexandru:2016gsd}.


\section{Exact ground-state energy}
We first consider the (Lorentzian) ${\mathbb C}P^1$ quantum 
mechanics described by the Lagrangian 
\beq
g^2 L \hs{-1} &=& \hs{-1} G \Big[ |\p_t \varphi|^2 - 
|m \varphi|^2+ i \bar \psi \D_t \psi \Big] 
- \epsilon \, \frac{\p^2 \mu}{\p \varphi \p \bar \varphi} \, \psi \bar \psi, 
\label{eq:L_{CP1}}
\eeq 
where $\varphi$ is the inhomogeneous coordinate, 
$G = \p_\varphi \p_{\bar \varphi} \log(1+|\varphi|^2)$ is the 
Fubini-Study metric, $\D_t = \p_t + \p_t \varphi \, \p_\varphi \log G$ 
is the pull back of the covariant derivative and 
$\mu= m|\varphi|^2/(1+|\varphi|^2)$ is the moment map 
associated with the $U(1)$ symmetry $\varphi \rightarrow 
e^{i \theta} \varphi$. 
The parameter $\epsilon$ is the boson-fermion coupling 
and the Lagrangian becomes supersymmetric at $\epsilon=1$.
Since the fermion number $F= G \psi \bar \psi$ commutes with 
the Hamiltonian, the Hilbert space can be decomposed into two 
subspaces with $F = 1$ and $F=0$. 
By projecting quantum states onto the subspace which contains 
the ground state ($F=1$), we obtain the bosonic 
Lagrangian 
\beq
L = {|\p_t \varphi|^2 \over{(g^2(1+|\varphi|^2)^2)}}-V\,,
\eeq
with 
the 
potential 
\beq
V = 
 \frac{1}{g^2} \frac{m^2 |\varphi|^2}{(1+|\varphi|^2)^2} 
- \epsilon m \frac{1-|\varphi|^2}{1+|\varphi|^2}.
\label{eq:potential}
\eeq
We note that $\theta(\equiv-2\arctan|\varphi|)=0,\pi$ are
global and metastable vacua respectively.

For $\epsilon = 1$, the ground state wave function $\Psi_0$ 
preserving the SUSY is given as a zero energy solution of the 
Schr\"odinger equation 
\beq
H_{\epsilon = 1} \Psi_0 = \left[ - g^2 (1+|\varphi|^2)^2 
\frac{\p}{\p \varphi} \frac{\p}{\p \bar \varphi} + V_{\epsilon = 1} 
\right] \Psi_0 = 0. 
\label{eq:H}
\eeq
It is exactly solved as 
\beq
\Psi_0 = \langle \varphi | 0 \rangle = \exp( - \mu/g^2)
\eeq
For $\epsilon \approx 1$, the leading order correction to the 
ground state wave function can be obtained by expanding the 
Schr\"odinger equation with respect to small 
$\delta \epsilon \equiv \epsilon-1$ as $\langle \varphi | \delta \Psi \rangle$.
Correspondingly, the ground state energy $E$ can also be expanded 
\begin{align}
E = \delta\epsilon\, E^{(1)} \, + \, \delta\epsilon^{2} \, 
E^{(2) }\, + \, \cdots \, . 
\label{eq:nearsusy}
\end{align}
These expansion coefficients can be determined by the standard 
Rayleigh-Schr\"odinger perturbation theory as 
\beq
E^{(1)} &=& \frac{\langle 0 |\delta H|0 \rangle}{\langle 0 |0 \rangle},
\\
E^{(2)} &=& -\frac{\langle \delta \Psi |H_{\epsilon=1}|\delta \Psi \rangle}{\langle 0 |0 \rangle}, \cdots,
\eeq
with $\delta H ~=~ H - H_{\epsilon=1}$. 
We find that these coefficients $E^{(i)}$ are real without 
imaginary ambiguities and can be expanded in absolutely 
convergent power series with respect to the nonperturbative 
exponential $\exp(-2m/g^2)$ 
\beq
E^{(i)} = \sum_{p=0}^\infty E_p^{(i)} 
\exp 
( -{2pm}/{g^2} 
), 
\eeq
where the zero-th term $E_0^{(i)}$ corresponds to the perturbative 
contributions on the trivial vacuum (perturbative vacuum). 
The coefficients of $E^{(1)}$ 
\cite{Fujimori:2016ljw} are
\beq
E_{0}^{(1)} = - m + g^2 , \hs{5} 
E_p^{(1)} = - 2m, \; (p\ge 1).
\label{eq:E1p}
\eeq
If the coefficients of $E^{(2)}$ are expanded in powers of $g^2$, 
they give factorially divergent asymptotic series, which can 
be Borel-resummed. Hence we rewrite the coefficient in the 
form of the Borel transform (See Appendix.~A for the details of calculations.) as 
\beq
E_{0}^{(2)} &=& g^{2}+ 2m \int_0^\infty dt 
\frac{e^{-t}}{t-\frac{2m}{g^2 \pm i0}}, 
\label{eq:E20}
\\
E^{(2)}_p &=& 2m \int_0^\infty dt \, e^{-t} \left\{ 
\frac{(p+1)^2}{t-\frac{2m}{g^2\pm i0}} + \frac{(p-1)^2}{t+\frac{2m}{g^2}} 
\right\} \notag \\
&+& 4 m p^2 \left( \gamma + \log \frac{2m}{g^2} \pm \frac{\pi i}{2} \right), 
\; (p\ge 1).
\label{eq:E2p}
\eeq
Note that the imaginary ambiguities associated to the Borel 
resummation is manifest in the first term of $E_p^{(2)}$ 
with $g^2\pm i0$, which is compensated by the imaginary 
part $\pm i\pi/2$ in the last term of $E_{p+1}^{(2)}$, 
reproducing the original real $E^{(2)}$ precisely. 
In the present case, we have only poles in the Borel plane while cuts are expected for general cases. We also note that in \cite{Dunne:2016jsr} the perturbation series on 0-bion background including the level number information has been shown to give all p-bion contributions.

We can now recognize the full resurgence structure to all orders 
of nonperturbative exponential: imaginary ambiguity of the 
non-Borel summable divergent perturbation series on the 
$p$-bion background in the first term of $E_p^{(2)}$ 
is cancelled by the imaginary ambiguity of the classical 
contribution of $(p+1)$-bion contribution in the last term of 
$E_{p+1}^{(2)}$. 
We note the absence of powers of $g^2$ in the imaginary ambiguity, 
which will allow us to recover non-Borel summable perturbation series 
on the $p$-bion background completely from the $(p+1)$-bion contribution
through the dispersion relation, 
without computing perturbative corrections around the multi-bion background explicitly. 
Moreover, if we observe that $E^{(2)}/m$ is an even function of 
$m/g^2$, we can also understand the presence of Borel-summable 
part (second term of the first line in Eq.(\ref{eq:E2p})). 
Thus all the terms can now be reproduced through resurgence 
relation and the sign change of $m/g^2$, 
if we can compute all the semi-classical $p$-bion contributions.


\section{Multi-bion solutions}
Nonperturbative contributions to the ground 
state energy come from the saddle points of the path integral 
$ Z = \int \D\varphi \D \tilde \varphi \, e^{-S_E} \,\sim\, 
e^{-\beta E}$ (for large $\beta$), 
where we have complexified the degrees of freedom by regarding 
$\varphi \ \equiv \varphi_R^\C+i\varphi_I^\C$ and 
$\tilde \varphi \ \equiv \varphi_R^\C-i\varphi_I^\C$ 
as independent holomorphic variables, and imposed the periodic 
boundary condition $\varphi(\tau+\beta)=\varphi(\tau)$ and 
for $\tilde\varphi$. 
The Euclidean action 
\beq
S_E =  \int_0^\beta d\tau [ \p_\tau \varphi \p_\tau \tilde \varphi/(g^{2}(1 + \varphi \tilde \varphi)^2) + V(\varphi \tilde \varphi) ],
\eeq
has two conserved Noether 
charges associated with the complexification of the Euclidean 
time translation $\tau \rightarrow \tau + a$ and 
the phase rotation $(\varphi, \tilde \varphi) \rightarrow 
(e^{ib} \varphi, e^{-ib} \tilde \varphi)$ 
($a,b \in {\mathbb C}$). 
Using the corresponding conservation laws, we can obtain 
the following solution of the equation of motion with nontrivial 
contribution in a $\beta\to\infty$ limit,
\beq
\varphi = e^{i \phi_c} \frac{f(\tau-\tau_c)}{\sin \alpha}, \hs{5}
\tilde \varphi = e^{-i \phi_c} \frac{f(\tau-\tau_c)}{\sin \alpha}, 
\label{eq:sol}
\eeq 
where $(\tau_c,\phi_c)$ are complex moduli parameters associated with 
the symmetry and $f(\tau)$ is the elliptic function 
\beq
f(\tau) = {\rm cs} (\Omega \tau, k) \equiv  
{\rm cn} (\Omega \tau, k)/ {\rm sn} (\Omega \tau, k), 
\eeq
which satisfies the differential equation 
\beq
(\p_\tau f)^2 = \Omega^2 (f^2+1)(f^2+1-k^2)\,.
\eeq
Solutions are characterized by two integers $(p,q)$ 
for the period 
\beq
\beta={(2pK+4iqK')\over{\Omega}}
\eeq
with $2K(k)$ and $4iK'$ ($K'\equiv K(\sqrt{1-k^2})$) as the period of 
the doubly periodic function cs. 
The parameters $(\alpha, \Omega, k)$ are given in terms of the 
period $\beta$, and their asymptotic forms for large $\beta$ 
(See Appendix.~B for the details of calculations.)
are given by 
\beq
&k \approx 1 - 8 \, e^{ - \frac{\omega \beta - 2 \pi i q}{p}}, \hs{3}
\Omega \approx \omega \left( 1 + 8 \frac{\omega^2+m^2}{\omega^2-m^2} 
e^{ - \frac{\omega \beta - 2 \pi i q}{p}} \right), & \notag \\
&\cos \alpha \approx \frac{m}{\omega} \left( 1 - \frac{8m^2}{\omega^2-m^2} 
e^{ - \frac{\omega \beta - 2 \pi i q}{p}} \right) , &
\label{eq:parameters}
\eeq
where $\omega = m \sqrt{1+ 2 \epsilon g^2/m}$ and 
$(p,q)$ are arbitrary integers such that $0 \leq q < p$. 
The asymptotic value of the action for the $(p,q)$ solution  
is given by 
\beq
S \approx p S_{\rm bion} + 2 \pi i \epsilon l , \hs{3} 
S_{\rm bion} = \frac{2m}{g^2} + 2 \epsilon \log 
\frac{\omega + m}{\omega - m},
\label{eq:action}
\eeq
where we have ignored the vacuum value of the action. 
\begin{figure}[t]
\includegraphics[width=90mm]{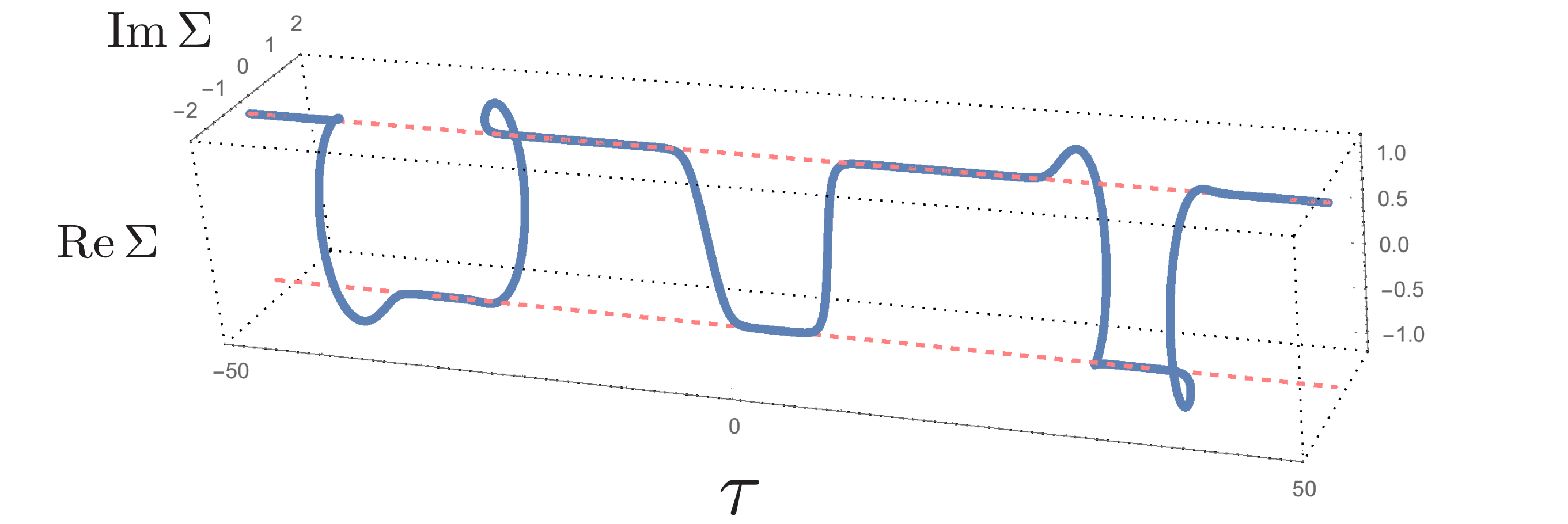}
\caption{Multi-bion solution: kink profile of 
$\Sigma(\tau)=(1-\varphi \tilde \varphi)/(1+\varphi \tilde \varphi)$ 
for $(p,q)=(3,1)$, $\epsilon=1$, $m=1$, $g = 1/200$, 
$\beta=100$ and $\tau_{c}=0$. $\Sigma=\pm1$ (dashed lines) correspond to
north and south poles (global and local minima) of $\C P^1$.}
\label{fig:multi_bion}
\end{figure}
The imaginary part $2 \pi i \epsilon l$ is related to the 
so-called hidden topological angle \cite{Behtash:2015kna} 
and the integer $l$ is zero or the greatest common divisor of 
$p$ and $2q$ depending on the value of ${\rm Im} \, \tau_c$.
We see that the integer $p$ is the number of bions, and that 
the $n$-th kink and antikink are located at $\tau_n^+$ and 
$\tau_n^-$,
with 
\beq
\tau_n^\pm = \tau_c + \frac{n-1}{\omega p} (\omega \beta - 2\pi i q) 
\pm \frac{1}{2\omega}\log 
{\frac{4\omega^2}{\omega^2-m^2}}.
\label{eq:QM_saddle}
\eeq
There are $p$ bions (pairs of kink-antikink) equally spaced on $S^1$, 
In Fig.~\ref{fig:multi_bion}, we depict the profile of the 
complexified height function 
\beq
\Sigma = { (1-\varphi \tilde \varphi)\over{(1+\varphi \tilde \varphi)}}\,,
\eeq
of the $(p,q)=(3,1)$ solution.
It illustrates that general $(p,q)$ solutions are intrinsically complex, 
and are not a mere repetition of single (real or complex) bions. 
In Fig.~\ref{fig:multi_bion2} we depict other solutions $(p,q)=(2,0)$ 
and $(p,q)=(2,1)$ in terms of $\theta=-2 \arctan|\varphi|$, which 
visualizes patterns of transition between the (metastable) vacua.
Although our solutions are not solutions of the SUSY theory with fermions, they are composite configurations of instantons and anti-instantons which are typically non-BPS. This fact implies that the non-BPS configurations play a vital role in the semi-classics in the path integral formalism of quantum theories.


\section{Multi-bion contributions}
The contributions from the $p$-bion solutions can 
be calculated by performing the Lefschetz thimble integral
associated with the saddle points. 
In the weak coupling limit $g \rightarrow 0$, 
we can use the Gaussian approximation 
for the fluctuation modes from the saddle points
except the nearly massless modes 
parameterized by the quasi moduli parameters $(\tau_i, \phi_i)$. 
Thus, we can simplify the Lefschetz thimble analysis
by reducing the degrees of freedom onto the quasi moduli space.

\begin{figure}[t]
\includegraphics[width=60mm]{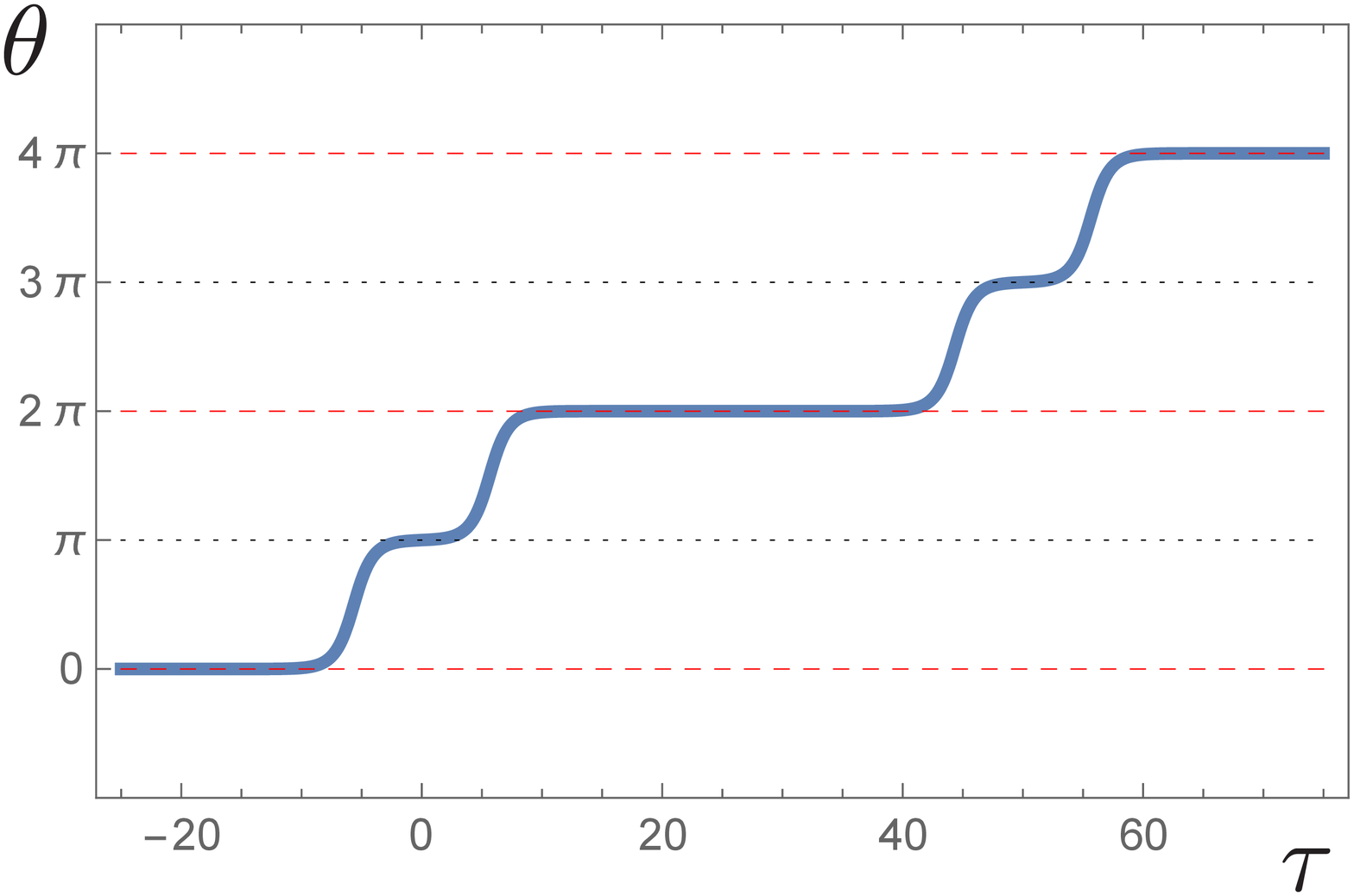}
\includegraphics[width=60mm]{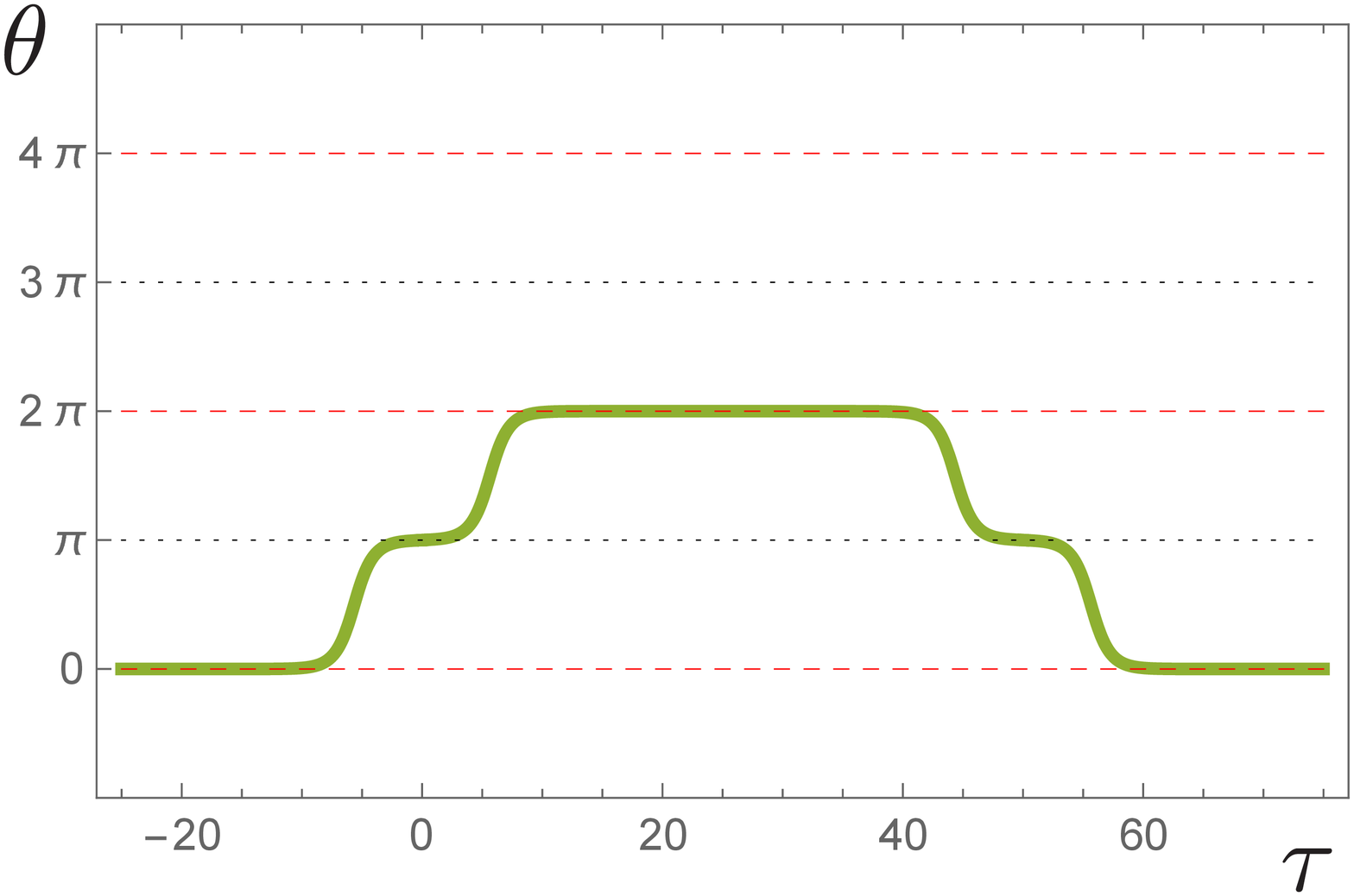}
\caption{Multi-bion solution: $\theta=-2 \arctan|\varphi|$
for $(p,q)=(2,0)$ (top) and for $(p,q)=(2,1)$ (bottom). 
The other parameters are the same as those in Fig.~\ref{fig:multi_bion}. 
$\theta=0,2\pi,...$ and $\theta=\pi,3\pi,...$ correspond to
north and south poles of $\C P^1$.}
\label{fig:multi_bion2}
\end{figure}

The leading order contributions come from 
the region around the saddle points, 
where all the kinks are well-separated in the weak coupling limit. 
Therefore, the effective potential can be approximated by that 
for well-separated kinks
\beq
S_E ~\rightarrow~ V_{\rm eff} = - m \epsilon \beta + \sum_{i=1}^{2p} ( \frac{m}{g^2} + V_{i} )\,,
\eeq
where $V_{i}$ is the asymptotic interaction potential between 
neighboring kink-antikink pair \cite{Misumi:2014jua}
\beq
{V_{i}\over{m}} = \epsilon_i (\tau_i - \tau_{i-1}) - \frac{4}{g^2} 
e^{-m(\tau_i-\tau_{i-1})} \cos(\phi_i-\phi_{i-1}),
\eeq
with $\tau_{2n-1} = \tau_i^-$, $\tau_{2n} = \tau_i^+$, 
$\tau_0 = \tau_{2p}-\beta$, $\phi_0=\phi_{2p}~({\rm mod} \, 2\pi)$, 
$\epsilon_{2n-1} = 0$ and $\epsilon_{2n} = 2\epsilon$. 
We find that the saddle points of $V_{\rm eff}$ are consistent 
with $\tau_{n}^{\pm}$ in Eq.~(\ref{eq:QM_saddle}) for large 
$\beta$ and small $g$ . 
We introduce a Lagrange multiplier $\sigma$ to impose the 
periodicity as 
\beq
2\pi \delta\left(\sum_{i} \tau_{i} - \beta\right)=m\int d\sigma\,{\rm exp}\left[im\sigma(\sum_{i} \tau_{i} -\beta)\right]\,.
\eeq
By generalizing the Lefschetz thimble analysis in 
\cite{Fujimori:2016ljw} to the multi-bion contribution 
\beq
Z_p \propto \int \prod_{i=1}^{2p} d \tau_i d \phi_i , \exp (-V_{\rm eff})\,,
\eeq
we obtain the following $p$-bion contribution to the partition 
function (See Appendix.~C for the details of calculations.) 
\beq
\frac{Z_p}{Z_0} \! \! &\approx& \! \! - \frac{2 i m \beta}{p} 
e^{-\frac{2pm}{g^2}} \, \underset{\sigma = 0}{\rm Res} 
\left[ e^{- i m \beta \sigma} \prod_{i=1}^{2p} I_i \right],
\label{eq:Zp}
\eeq
with 
\beq
I_i ~=~ 
\frac{2m}{g^2} \left( \frac{2m}{g^2} e^{\pm \frac{\pi i}{2}} \right)^{i \sigma- \epsilon_i} 
\frac{\displaystyle \Gamma \left( (\epsilon_i - i\sigma )/2 \right)}{\displaystyle \Gamma \left( 1 - (\epsilon_i - i \sigma) / 2 \right)}. 
\eeq
The sign $\pm$ is associated with ${\rm arg}[g^{2}]=\pm0$.
This gives a polynomial of $\beta$, whose leading term is of order $\beta^p$ 
\beq
\frac{Z_p}{Z_0} \! \! &\approx& \! \! \frac{1}{p!} \left[ 
\frac{2 m \beta \Gamma(\epsilon)}{\Gamma(1-\epsilon)}
e^{-\frac{2m}{g^2} \mp \pi i \epsilon} 
\left( \frac{2m}{g^2} \right)^{2(1- \epsilon)} 
 \right]^p,
\label{eq:p-bion_leading}
\eeq
consistent with the dilute gas approximation: 
$Z_p/Z_0 = (Z_1/Z_0)^p/ p! + \mathcal O(\beta^{p-1})$. 
From the $p$-bion contribution \eqref{eq:Zp} and the 
perturbative contribution ($p=0$), 
the ground state energy $E = - \lim_{\beta \rightarrow \infty} 
\frac{1}{\beta} \log Z$ can be obtained as 
\beq
E ~=~ E_{0
} - \lim_{\beta \rightarrow \infty} \frac{1}{\beta} 
\log \left( 1 + \sum_{p=1}^\infty \frac{Z_p}{Z_0} \right). 
\eeq
By taking the logarithm, contributions of high powers of 
$\beta$ such as $\beta^{p}$ for $p>1$ should be cancelled, 
and the ground state energy is obtained from the remaining 
contributions of order $\beta$. 
Fortunately, most of these contributions with high powers of 
$\beta$ disappear near SUSY case thanks to the zero in 
$1/\Gamma(1- (\epsilon_i - i \sigma)/2)$. 
As a result, we find that the first derivative is proportional 
to $\beta$ and gives the near-SUSY ground state energy $E^{(1)}$ 
\begin{equation}
E^{(1)}_p=-e^{2pm\over{g^{2}}}\lim_{\epsilon\to 1}\lim_{\beta\to\infty}
\frac{1}{\beta}
\frac{\partial}{\partial \epsilon}\frac{Z_p}{Z_0}
=-
2m, 
\end{equation}
verifying the exact result (\ref{eq:E1p}). 
The second derivative in $\epsilon$ turn out to be quadratic 
in $\beta$, and 
\beq
E^{(2)}_p =-\frac{e^{2pm\over{g^2}}}{2}  \lim_{\epsilon\to 1} \lim_{\beta\to\infty}\frac{1}{\beta} \left[ \p_{\epsilon}^2 \frac{Z_p}{Z_0} - \sum_{i=1}^{p-1} \p_{\epsilon} \frac{Z_{p-i}}{Z_0} \p_{\epsilon} \frac{Z_{i}}{Z_0} \right]\,,
\nonumber\\
\eeq
is calculated as
\beq
E^{(2)}_p = 4mp^{2} \left( \gamma +\log {2m\over{g^{2}}} \pm {\pi i \over{2}}  \right),
\eeq 
in complete agreement with the exact result (\ref{eq:E2p}). 
We have obtained the classical contributions to all orders of 
multi-bions, which provides all terms needed for the full 
resurgence structure of our model, 
although it is difficult to check the divergent perturbation series 
on $p$-bion background directly, except for the trivial vacuum ($p=0$).

\section{Perturbation series on trivial vacuum}
We obtain the perturbation series on the trivial background ($p=0$) 
by using the Bender-Wu method\cite{BenderWu,Sulejmanpasic:2016fwr}. 
We first expand the energy and the wave function as 
\beq
E = m \sum_{l} A_l \eta^{2l}, \,\,\,\,\,\,\,\,
\Psi = \exp (- x^2) \sum_{l,k} B_{l,k} \eta^{2l} x^{2k}\,,
\eeq
with $|\varphi| = \eta x$ and $\eta^2 = g^2 /m$. 
Then, the Schr\"odinger equation reduces to a (Bender-Wu) 
recursive equation for $A_l$ and $B_{l,k}$, 
which gives the leading asymptotic behavior 
(See Appendix.~D for the details of calculations.)
as 
\beq
A_l \sim - \frac{\Gamma(l+2(1-\epsilon))}{2^{l-1} 
\Gamma(1-\epsilon)^2}, ~~~\mbox{(for large $l$)}. 
\label{eq:Al}
\eeq
Since the coefficients $A_l$ grow factorially for large $l$, 
we obtain the perturbative part of the ground state energy 
by using the Borel resummation 
\beq
E_{0} \sim \frac{2m}{\Gamma(1-\epsilon)^2} \int_0^\infty dt \, e^{-t} \ t^{2(1-\epsilon)}(t-\frac{2m}{g^2})^{-1}\,.
\eeq
The Borel resummation gives a finite result with the 
imaginary ambiguity 
\beq
{\rm Im} \, E_{0} = \mp \frac{2\pi m}{\Gamma(1-\epsilon)^2} \left( \frac{g^2}{2m} \right)^{2(\epsilon-1)} e^{-\frac{2m}{g^2}},
\eeq
with $-$ ($+$) in the right hand side for ${\rm Im} g^{2}= +0$ ($-0$).
This imaginary ambiguity of the perturbation series in the trivial 
vacuum ($p=0$) cancels that of the single bion sector 
(\eqref{eq:p-bion_leading} with $p=1$). 
Therefore, combining these two contributions gives unambiguous 
real result. 
This result verifies the resurgence for arbitrary values of 
$\epsilon$ including the non-SUSY case explicitly, although 
only to the leading order of nonperturbative exponential.

For the near-SUSY case, we can obtain the perturbation series 
on the trivial vacuum exactly to all orders in $g^2$, 
by exactly solving the Bender-Wu recursion relation 
to the second order of $\delta \epsilon$ as 
\beq
E_{0
} = (g^2-m) \delta \epsilon - 2m 
\sum_{l=2}^\infty \Gamma(l) \left(\frac{g^2}{2m}\right)^{l} \delta \epsilon^2 + \cdots.
\eeq 
This agrees completely with the exact results $E_{0}^{(1)}$ in Eq.~\eqref{eq:E1p} 
and $E_{0}^{(2)}$ in Eq.~\eqref{eq:E20} after Borel-resummation.

\section{Summary and Discussion}
In conclusion, 

(i) We have derived the exact expansion coefficients of the 
ground state energy to the second order of the SUSY breaking 
deformation parameter $\delta \epsilon$. 
The result shows a resurgent trans-series structure to all 
order of nonperturbative exponential. 

(ii) We have derived nonperturbative multi-bion contributions 
with imaginary ambiguities in the weak coupling limit and 
found that they agree with the corresponding parts in the 
exact result.

(iii) At least for near-SUSY ${\mathbb C} P^1$ QM, 
by assuming the cancellation of imaginary ambiguities (resurgence structure) 
and an even function of $m/g^2$, we have recovered the entire 
trans-series which agrees with the exact result of the near-SUSY.  

(iv) With the Bender-Wu recursion relation, we have 
obtained the perturbation series on $0$-bion vacuum to all orders, 
which gives an imaginary ambiguity when Borel-resummed, 
and have verified the cancellation with that of single bion 
sector for general deformation parameter $\epsilon$ including 
non-SUSY case.

The exact result in Eq.(\ref{eq:E2p}) shows that the imaginary 
ambiguities have no $g^2$ corrections in ${\mathbb C}P^1$ QM. 
This fact enabled us to recover the entire trans-series from 
the semi-classical multi-bion contributions only. 
In other models such as sine-Gordon QM, 
imaginary ambiguities from the multi-bion contribution have 
perturbative corrections in powers of $g^2$ \cite{Misumi:2015dua}. 
Then these perturbative corrections are needed in order to 
recover the full resurgent trans-series.

The same resurgence structure exists in ${\mathbb C} P^{N-1}$ 
models with $N > 2$. 
Similarly to $N=2$, we obtain $O(\delta\epsilon^{2})$ 
perturbative contribution with the imaginary ambiguity 
\beq
{\rm Im}E^{(2)}_{0}=\mp{N^{2}\pi\over{2}}\sum_{i=1}^{N-1}m_{i}A_{i}e^{-{2m_{i}\over g^{2}}}\,,
\eeq
where $A_{i}= \prod_{j=1, j\not= i}^{N-1}{m_{j}\over{m_{j}-m_{i}}}$ 
and the mass parameters $m_{i}$ are reduced from the 2D 
${\mathbb C} P^{N-1}$ model with twisted boundary conditions. 
We also calculate the $O(\delta\epsilon^{2})$ single-bion contribution
\beq
E^{(2)}_{1}=\sum_{i=1}^{N-1}N^{2}m_{i}A_{i}e^{-{2m_{i}\over{g^{2}}}}
\left(\gamma
+\log{2m_{i}\over{g^{2}}}\pm {\pi i\over2}\right).
\eeq
The imaginary ambiguities cancel between them.
As for convergence of $\delta\epsilon$ expansion,
we observe that each of the $p$-bion semiclassical contributions
has a convergent expansion for any $p$.

Focusing on the near-SUSY regime can be extended 
to the solvable models including localizable SUSY theories 
\cite{Aniceto:2014hoa, Honda:2016mvg} and quasi-solvable models 
\cite{Kozcaz:2016wvy} by softly breaking the solvable condition 
and expanding the physical quantities with respect to the deformation 
parameter. 
It is because these models have a similar resurgence property to the present $\C P^{1}$ model, where the resurgence structure becomes trivial without cancellation of imaginary ambiguity at localization-applicable or quasi-exactly-solvable regimes.
We also notice that the localization technique is applicable 
in ${\mathbb C}P^{N-1}$ QM to compute the first order ground 
state energy $E^{(1)}$ but not the second order. 
Recent results on volume independence \cite{Sulejmanpasic:2016llc} should be useful 
in extending our study to QFT, which may also require more 
refined thimble analysis as has been studied intensively
\cite{Cristoforetti:2013wha, Fujii:2013sra, Tanizaki:2014tua, Tanizaki:2014xba, Alexandru:2016gsd}.

Regarding non-SUSY gauge theories, complex instanton solutions were discussed in gauge theories with complexified gauge groups decades ago \cite{Dolan:1977hs,Burns:1983us}. It would be of importance to discuss contributions from these complex solutions in terms of resurgence theory.


\begin{acknowledgments}
The authors are grateful to the organizers and participants 
of ``Resurgence in Gauge and String Theories 2016" at IST, 
Lisbon and ``Resurgence at Kavli IPMU" at IPMU, U. of Tokyo 
for giving them a chance to deepen their ideas. 
This work is supported by 
the Ministry of Education, Culture, Sports, Science,
and Technology(MEXT)-Supported Program for the Strategic Research Foundation
at Private Universities ``Topological Science" (Grant No. S1511006).
This work is also supported in part 
by  the Japan Society for the 
Promotion of Science (JSPS) 
Grant-in-Aid for Scientific Research
(KAKENHI) Grant Numbers 
(16K17677 (T.\ M.), 16H03984 (M.\ N.) and 
25400241 (N.\ S.)).
The work of M.N. is also supported in part 
by a Grant-in-Aid for Scientific Research on Innovative Areas
``Topological Materials Science"
(KAKENHI Grant No. 15H05855) and 
``Nuclear Matter in neutron Stars investigated by experiments and
astronomical observations"
(KAKENHI Grant No. 15H00841) 
from MEXT of Japan.
\end{acknowledgments}

\appendix


\section{Exact ground-state energy}
In this section we show details of calculations in the part of ``Exact ground-state energy".
The leading order correction to the ground-state wave function and energy for ${\mathbb C}P^{1}$ quantum mechanics in Eqs.~(\ref{eq:L_{CP1}})(\ref{eq:potential}) can be obtained by solving the $\mathcal O(\delta \epsilon)$ 
part of the Schr\"odinger equation
\beq
H_{\epsilon=1} | \delta \Psi \rangle  &=& \left( E^{(1)} + m \frac{1-|\varphi|^2}{1+|\varphi|^2} \right) |0 \rangle, \\
E^{(1)} &=& g^2 - m \coth \frac{m}{g^2} 
 \notag \\
&=& -m + g^{2} \,-\, \sum_{p=1}^{\infty} 2m e^{-{2pm\over{g^{2}}}}.
\eeq
From this expanded form, 
we can read the expansion coefficients in Eq.\,\eqref{eq:E1p}. 
The above differential equation can be exactly solved as
\beq
\langle \varphi | \delta \Psi \rangle &=& e^{-\frac{\mu}{g^2}} \int_0^\mu \frac{d \mu'}{\mu'(\mu'-m)} \left( \mu' - m \frac{1-e^{\frac{2\mu'}{g^2}}}{1-e^{\frac{2m}{g^2}}} \right). \notag
\\
\eeq
Then we find the second-order correction to the ground-state energy as
\beq
E^{(2)} ~&=&~ -\frac{\langle \delta \Psi| H_{\epsilon=1} | \delta \Psi \rangle}{\langle 0 | 0 \rangle} 
\\ \notag
~&=&~ g^2 - 2 m \frac{\cosh \frac{m}{g^2}}{\sinh^3 \frac{m}{g^2}} \int_0^m d \mu \frac{\sinh^2 \frac{\mu}{g^2}}{\mu}. 
\eeq
Using the hyperbolic cosine integral ${\rm Chi}(z)$ defined by
\beq
{\rm Chi} (z) = \gamma + \log z - \int_0^z \frac{dt}{t} (1-\cosh t),
\eeq
we can rewrite $E^{(2)}$ as
\beq
E^{(2)} = g^2 - m \frac{\cosh \frac{m}{g^2}}{\sinh^3 \frac{m}{g^2}} \left[ {\rm Chi} \left(\frac{2m}{g^2}\right) - \gamma - \log \frac{2m}{g^2} \right]. \notag \\
\eeq
By using the relation
\beq
&&{\rm Chi} \left( \frac{2m}{g^2} \right) 
\\ \notag
&&~=~ - \frac{1}{2} \int_0^\infty dt \, e^{-t} \left( \frac{e^{\frac{2m}{g^2}}}{t-\frac{2m}{g^2 \pm i0}} + \frac{e^{-\frac{2m}{g^2}}}{t+\frac{2m}{g^2}} \right) \mp \frac{\pi i}{2}, 
\eeq
$E^{(2)}$ can be expanded as
\beq
E^{(2)} &=& g^2 + 2 m \int_0^\infty dt \, e^{-t} \frac{1}{t
-\frac{2m}{g^2 \pm i0}} 
\notag \\
&+& 4m \sum_{p=1}^\infty e^{-\frac{2pm}{g^2}} \Bigg[ p^2 \left( \gamma + \log \frac{2m}{g^2} \pm \frac{\pi i}{2} \right) \notag \\
&+& \frac{1}{2} \int_0^\infty dt \, e^{-t} \left( \frac{(p+1)^2}{t-\frac{2m}{g^2 \pm i0}} + \frac{(p-1)^2}{t+\frac{2m}{g^2}} \right) \Bigg]. \notag \\
\eeq
From this expanded form, 
we can read the expansion coefficients Eq.\,\eqref{eq:E20} and Eq.\,\eqref{eq:E2p}.


\section{Multi-bion solutions}
In this section we summarize basic properties of the multi-bion solution Eq.\,\eqref{eq:sol}
\beq
&&\varphi = e^{i \phi_c} \frac{f(\tau-\tau_c)}{\sin \alpha}, \,\,\,\,\,
\tilde \varphi = e^{-i \phi_c} \frac{f(\tau-\tau_c)}{\sin \alpha}, 
\\ \notag
&&f(\tau) = {\rm cs}(\Omega\tau,k),
\eeq
where the parameters are related as
\beq
&&k^2 ~=~ 1 - \tan^2 \! \alpha \left( \cos^2 \! \alpha - \frac{m^2}{\Omega^2} \right), \hs{10}
\\ \notag
&&\Omega ~=~ \omega \, \sqrt{1 -  \left( 1 + \sec^2 \! \alpha \right) \left( 1 - \frac{m^2}{\omega^2} \sec^2 \! \alpha \right)},
\label{eq:parameters_SM}
\eeq
with
\beq
\omega = m \sqrt{1+ \frac{2 \epsilon g^2}{m}}.
\eeq
This is a periodic solution, whose period is given by
\beq
\beta ~=~ \oint \frac{df}{\p_\tau f} ~=~ \frac{1}{\Omega} \oint \frac{df}{\sqrt{(f^2+1)(f^2+1-k^2)}}, 
\eeq
where we have used the relation $(\p_\tau f)^2 = \Omega^2 (f^2+1)(f^2+1-k^2)$. 
There are four branch points corresponding to the turning points $(\p_\tau \varphi = \p_\tau \tilde \varphi = 0)$
\beq
f = \pm i, ~ \pm i \sqrt{1-k^2}. 
\eeq
Let us introduce two branch cuts on the lines from $\pm i$ to $\pm i\sqrt{1-k^2}$ on the complex $f$-plane.
Let $C_A$ be the cycle from ${\rm Re} \, f = -\infty$ to ${\rm Re} \, f = \infty$ which does not pass through the branch cuts
and $C_B$ be the cycle surrounding the two branch points $\pm i \sqrt{1-k^2}$. 
Their periods are
\beq
\beta_A ~=~ \frac{2K(k)}{\Omega} , \hs{10} \beta_B ~=~ \frac{4 i K(\sqrt{1-k^2})}{\Omega},
\eeq 
where $K(k)=F(\pi/2,k)$ is the complete elliptic integral of the first kind
\beq
F(x,k) = \int_0^x \frac{d \theta}{\sqrt{1-k^2 \sin^2 \theta}}. 
\eeq
The period of the solution winding the cycle $p C_A + q C_B~(p,q \in \Z)$ is given by 
\beq
\beta = \frac{2 p K(k) + 4 q i K(\sqrt{1-k^2})}{\Omega}, \hs{10} p,q \in \Z. 
\eeq
Solving this equation and Eq.\,\eqref{eq:parameters_SM},
we can determine the parameters $(\alpha, \Omega, k)$ for each 
pair of integers $(p,q)$. 
The $\beta\to\infty$ limit of $(p,q)=(1,0)$ solution is given 
by the known one-bion solution for infinite time interval 
with 
\beq
E=m\epsilon, \hs{5} k=1, \hs{5} \cos \alpha = \frac{m}{\omega}, 
\hs{5} \Omega = \omega. 
\eeq
We need the $\beta\to\infty$ limit 
keeping $(p,q)$ fixed. 
Expanding the period with respect to $\delta k = k - 1$, 
we find that
\beq
\beta = \frac{1}{\omega} \left[ - p \log \left( -\frac{\delta k}{8} \right) + 2 \pi i q \right] + \mathcal O (\delta k \log \delta k).
\label{eq:period}
\eeq
Therefore, the asymptotic form of $\delta k$ for large $\beta$ is
\beq
\delta k \approx - 8 \, e^{ - \frac{\omega \beta - 2 \pi i q}{p}}.
\eeq
We can also show that the asymptotic forms of other parameters are 
\beq
&&\delta \alpha \approx \left( \frac{4m^2}{\omega^2-m^2} \right)^\frac{3}{2} e^{ - \frac{\omega \beta - 2 \pi i q}{p}}, \hs{3}
\\ \notag
&&\delta \Omega =  8 \omega \frac{\omega^2+m^2}{\omega^2-m^2} e^{ - \frac{\omega \beta - 2 \pi i q}{p}}, \hs{3}
\\ \notag
&&\delta E = \frac{8\omega^2}{g^2} \frac{\omega^2}{\omega^2-m^2} e^{ - \frac{\omega \beta - 2 \pi i q}{p}}. 
\eeq
We read Eq.~(\ref{eq:parameters}) from these equations.
Note that Eq.\,\eqref{eq:period} implies that the solution exists only for $0 \leq q < p$ in the large $\beta$ limit.

The action for this solution is given by
\beq
S_{\rm sol} = \int_0^\beta d\tau \, L_{\rm sol} =-m \epsilon \beta + \frac{\Omega}{g^2} \oint df \, X(f), 
\eeq
where
The function $X(f)$ can be written as
\beq
\notag
X &=& - \frac{\p}{\p f} \Big[ \frac{\omega^2 - m^2}{\Omega^2} \left\{ \frac{F(x,k)}{\cos^2 \! \alpha} - \tan^2 \! \alpha \, \Pi(\cos^2 \! \alpha, x, k) \right\} 
\\ 
&+&E(x,k) - \sqrt{\frac{f^2+1-k^2}{f^2+1}} \frac{f \cos^2 \! \alpha}{f^2 + \sin^2 \! \alpha} \Big]
\\ \notag
&=& - i \frac{\p}{\p f} \Big[ \frac{\omega^2 - m^2}
{\Omega^2} \left\{ F(y,k') - \Pi \left( \frac{1-k^2}{\sin^2 \! \alpha},y, k' \right) \right\} 
\\ \notag
&+& F(y,k') - E(y,k') + i \frac{f\sqrt{(f^2+1)(f^2+1-k^2)}}{f^2 + \sin^2 \! \alpha} \Big], \notag
\eeq
with $x= \arcsin \sqrt{\frac{1}{f^2+1}}$, $y= \arcsin \sqrt{\frac{-f^2}{1-k^2}}$ and $k' = \sqrt{1-k^2}$. 
Then we obtain
\beq
X(f) = \frac{1}{\sqrt{(f^2+1)(f^2+1-k^2)}} \Big[ - \frac{1-k^2}{\sin^2 \! \alpha} 
\notag \\
+ (f^2+1)(f^2+1-k^2) \frac{2 \sin^2 \! \alpha}{(f^2 + \sin^2 \! \alpha)^2} \Big].
\eeq
There are contributions from $C_A$, $C_B$ and the poles 
at $f = \pm i \sin \alpha$ (more precisely, integration cycles should be defined on the torus with two punctures)
\beq
S = - m \epsilon \beta + p S_A + q S_B + 2 \pi i l S_{\rm res}, \hs{7} l \in \Z.
\eeq
Explicitly, $S_{\rm res} = \epsilon$ and $S_A$ and $S_B$ are given by
\beq
S_A = \frac{2\Omega}{g^2} \ \Bigg[ \frac{\omega^2 - m^2}{\Omega^2} \left\{ \frac{K(k)}{\cos^2 \! \alpha} - \tan^2 \! \alpha \, \Pi(\cos^2 \! \alpha, k) \right\} 
\notag \\
+ E(k) \Bigg], 
\notag \\
S_B = \frac{4i \Omega}{g^2} \Bigg[ \frac{\omega^2 - m^2}{\Omega^2} \left\{ K(k') - \Pi\left( \frac{1-k^2}{\sin^2 \! \alpha}, k' \right) \right\} 
\notag \\
+ K(k') - E(k') \Bigg],\notag \\
\eeq
where $E(k)=E(\frac{\pi}{2},k)$ and $\Pi(a, k)=\Pi(a,\frac{\pi}{2},k)$ are the complete elliptic integrals of the second and third kind
\beq
&&E(x,k) = \int_0^x d\theta \sqrt{1-k^2 \sin^2 \theta}, 
\\ \notag
&&\Pi(a,x,k) = \int_0^x \frac{d\theta}{\sqrt{1-k^2 \sin^2 \theta}} \frac{1}{1- a \sin^2 \theta},
\eeq
and $k'=\sqrt{1-k^2}$.
For large $\beta$,
\beq
S \approx - m \epsilon \beta + p \left[ \frac{2\omega}{g^2} + 2 \epsilon \log \frac{\omega+m}{\omega-m} \right] + 2 \pi i \epsilon l ,
\eeq
from which we read Eq.~(\ref{eq:action}).
This implies that the integer $p$ corresponds to the number of bions. 

Focusing on the region around 
\beq
\tau \approx \frac{n}{p} \beta \hs{5} (n = 0, 1, \cdots, p-1),
\eeq
we can approximate the solution for large $\beta$ as
\beq
h(\tau)
\approx
\sqrt{\frac{\omega^2}{\omega^2-m^2}} \left[ \sinh \left \{ \omega \left( \tau - \frac{n \beta}{p} \right) + \frac{2\pi i n q}{p} \right\} \right]^{-1}, \notag \\ 
\eeq
where we have used ${\rm cs}(x,1) = 1/ \sinh x$. 
Therefore, the solution in this region looks like the single bion configuration
\beq
&&\varphi \approx \left( e^{\omega (\tau - y_{n}^{+})} + e^{-\omega (\tau + \tilde y_{n}^{-})} \right)^{-1}, 
\\ \notag
&&\tilde \varphi \approx \left( e^{\omega (\tau - \tilde y_{n}^{+})} + e^{-\omega (\tau + y_{n}^{-})} \right)^{-1},
\eeq
with
\beq
\omega y_{n}^{\pm} &=& \omega \tau_c + i \phi_c + \frac{n}{p} \omega \beta - \frac{2\pi i n q}{p} \pm \log \sqrt{\frac{4\omega^2}{\omega^2-m^2}}  
\notag \\
&&\hs{40}({\rm mod} \ 2 \pi i) , \\
\omega {\tilde y}_{n}^{\pm} &=& \omega \tau_c - i \phi_c + \frac{n}{p} \omega \beta - \frac{2\pi i n q}{p} \pm \log \sqrt{\frac{4\omega^2}{\omega^2-m^2}} 
\notag \\
&&\hs{40}({\rm mod} \ 2 \pi i) . 
\eeq
From this asymptotic form, we can read off the positions 
$\tau_{n}^{\pm} = (y_{n-1}^{\pm} + \tilde y_{n-1}^{\pm})/2$
and phases $\phi_{n}^{\pm} = (y_{n-1}^{\pm} - \tilde y_{n-1}^{\pm})/2i$,
of the component kinks. 
The $n$-th kink ($+$) and antikink ($-$) locations Eq.~(\ref{eq:QM_saddle}) are given by  
\beq
\tau_n^\pm = \tau_c + \frac{n-1}{\omega p} (\omega \beta - 2\pi i q) \pm {1\over{2\omega}}\log \frac{4\omega^2}{\omega^2-m^2}.
\eeq

The poles of the Lagrangian are located at 
\beq
\omega \tau_{{\rm pole}, n\pm} 
&\approx& \omega \tau_c + \frac{n}{p} \omega \beta - \frac{2 \pi i n q}{p} 
\\ \notag
&\pm& {\rm arccosh} \sqrt{\frac{\omega^2}{\omega^2-m^2}} + \frac{\pi i}{2} ~~~~~~ ({\rm mod} \ \pi i). 
\eeq
These poles pass through the real $\tau$ axis for certain values of ${\rm Im} \, \tau_0$, 
at which the value of the action jumps discontinuously. 
When one of the poles, for example $\tau_{{\rm pole},n+}$, is on the real $\tau$ axis,
then $\tau_{{\rm pole},n'+}$ with $n' = n + k p/{\rm gcd}(p,2q)~(k=0,\cdots,{\rm gcd}(p,2q)-1)$ are also on the real $\tau$ axis, 
where ${\rm gcd}(p,2q)$ is the greatest common divisor of $p$ and $2q$.
Therefore, the discontinuity of the action when the poles pass through the real $\tau$ axis is
\beq
\Delta S = \pm 2 \pi i \epsilon \, {\rm gcd}(p,2q).
\eeq


\section{Multi-bion contributions}
In this section we explicitly evaluate the quasi moduli integral 
for the chain of $p$ kinks and $p$ anti-kinks alternately aligned on $S^1$ with period $\beta$. 
The effective potential consists of the nearest neighbor interactions 
\beq
V_{\rm eff} = - m \epsilon \beta + \sum_{i=1}^{2p} \left( \frac{m}{g^2} + V_i \right), 
\eeq
where $V_i$ is the interaction potential 
\beq
V_i \, = \,  m \epsilon_i (\tau_{i}-\tau_{i-1})- \frac{4m}{g^2} e^{-m (\tau_{i}-\tau_{i-1})} \cos(\phi_{i}-\phi_{i-1}) , \notag \\
\eeq
where $(\tau_i, \phi_i)$ are quasi moduli parameters corresponding to the position and phase of the $i$-th (anti)kink 
$(\tau_{0} = \tau_{2p} - \beta, ~ \phi_0 = \phi_{2p}~{\rm mod} \, 2 \pi)$ and 
\beq
\epsilon_i = \left\{ \begin{array}{ll} 
2 \epsilon & \mbox{for $i \in 2 \Z$} \\
0 & \mbox{for $i \in 2\Z +1$} \end{array} \right..
\eeq
It is convenient to redefine the relative quasi moduli parameters as
\beq
z_i = m (\tau_i - \tau_{i-1}) + i (\phi_i - \phi_{i-1}), 
\notag \\
\tilde z_i = m (\tau_i - \tau_{i-1}) - i (\phi_i - \phi_{i-1}). 
\eeq
Note that the imaginary parts of $z_i$ and $\tilde z_i$ are phases defined modulo $2 \pi$. 
The complex variables $z_i$ and $\tilde z_i$ are subject to the following constraints 
\beq
\sum_{i=1}^{2p} \frac{z_i + \tilde z_i}{2} = m \beta, \hs{5}
\sum_{i=1}^{2p} \frac{z_i - \tilde z_i}{2i} = 0 ~~~ ({\rm mod} \, 2 \pi),
\eeq
which are expressed by the integral forms of delta functions as functions of $\sigma$ and $s$
\beq
&&\delta \left( \sum_{i=1}^{2p} \frac{z_i+\tilde z_i}{2} - m \beta \right) 
\notag \\
&&\hs{20}= \int \frac{d\sigma}{2\pi} \, \exp \left[ i \sigma \left( \sum_{i=1}^{2p} \frac{z_i+\tilde z_i}{2} - m \beta \right) \right], \notag \\
&&\sum_{n=-\infty}^{\infty} \delta \left( \sum_{i=1}^{2p} \frac{z_i - \tilde z_i}{2i} - 2 \pi n \right) 
\notag \\
&&\hs{20}= \frac{1}{2\pi} \sum_{s=-\infty}^\infty \exp \left( s \frac{z_i - \tilde z_i}{2} \right).
\eeq
The saddle points $(q=0,1,\cdots, p-1)$ which give non-trivial contributions to the ground state energy are located at
\beq
z_i = \left\{ \begin{array}{ll} 
- \log \left( \sqrt{\left(\frac{\epsilon g^2}{4m}\right)^2 + e^{-\frac{m \beta - 2 \pi i q}{p}}} + \frac{\epsilon g^2}{4m} \right) 
\approx
 \\
\phantom{-} \log \frac{2m}{\epsilon g^2}  ~~~~(\mbox{for $i \in 2 \Z$}) \\ \\
- \log \left( \sqrt{\left(\frac{\epsilon g^2}{4m}\right)^2 + e^{-\frac{m \beta - 2 \pi i q}{p}}} - \frac{\epsilon g^2}{4m} \right) 
\approx 
\notag \\
- \log \frac{2m}{\epsilon g^2} + \frac{m \beta - 2 \pi i q}{p}  ~~~~(\mbox{for $i \in 2 \Z+1$})
\end{array} \right.. \notag \\
\eeq
We note that  the Lagrange multiplier $\sigma$ is expressed in terms of the other parameters on the saddle points.
This is consistent with the weak coupling limit $g^2 \rightarrow 0$ of Eq.\,\eqref{eq:QM_saddle}
with $z_{2n}/\omega = \tau_{n}^{+} - \tau_{n}^{-}$ and $z_{2n+1}/\omega = \tau_{n+1}^{-} - \tau_{n}^{+}$.

The $p$-bion contribution to the partition function is given by
\beq
\frac{Z_p}{Z_0} = \frac{1}{p} \int \prod_{i=1}^{2p} \left[ d \tau_{i} \wedge d\phi_{i} \, \frac{2m^2}{\pi g^2} \exp \left( - \frac{m}{g^2} - V_i \right) \right], 
\eeq
where the factor $\frac{2m^2}{\pi g^2}$ is the 1-loop determinant from the massive modes around each kink and the factor $1/p$ is inserted since the bions are indistinguishable.  
The integration measure can be rewritten as
\beq
&&\prod_{i=1}^{2p} d\tau_i \wedge d\phi_i = m d\tau_c \wedge d\phi_c \wedge \left( \prod_{i=1}^{2p} \frac{i}{2m} d z_i \wedge d \tilde z_i \right) \notag \\
&&  \times \delta \left( \sum_{i=1}^{2p} \frac{z_i+\tilde z_i}{2} - m \beta \right) \delta \left( \sum_{i=1}^{2p} \frac{z_i - \tilde z_i}{2i} - 2 \pi n \right), \notag \\
\eeq
where $\tau_c$ and $\phi_c$ are the overall moduli parameters. 
We can rewrite the $p$-bion contribution as
\beq
\frac{Z_p}{Z_0} = 
\frac{2\pi m \beta}{p} e^{-\frac{2pm}{g^2}} \sum_{s = -\infty}^{\infty} \frac{1}{4\pi^2} \int d\sigma \, e^{- i m \beta \sigma} \prod_{i=1}^{2p} I_i , \notag \\
\eeq
where
\beq
I_i = \frac{i m}{\pi g^2} \int d z_{i} \wedge d \tilde z_i \, \exp (- \mathcal V_i) \, \exp( - \tilde{\mathcal V}_i),
\eeq
with
\beq
\mathcal V_i = - \frac{2m}{g^2} e^{-z_i} + \frac{1}{2} (\epsilon_i - s - i \sigma) z_i , 
\notag \\
\tilde{\mathcal V}_i = - \frac{2m}{g^2} e^{-\tilde z_i} + \frac{1}{2} (\epsilon_i + s - i \sigma) \tilde z_i.
\eeq
We can show that the $p$-bion contribution satisfies the following differential equation 
\beq
&&\left[ \prod_{i=1}^{2p} \frac{s^2 - (\epsilon_i - i \hat \sigma)^2}{4} - \left( \frac{2m}{g^2} \right)^{4p} e^{-2 m \beta} \right] \left( \frac{1}{\beta} \frac{Z_p}{Z_0} \right) = 0, 
\notag \\
&&\hs{30}\hat \sigma = \frac{i}{m} \frac{\p}{\p \beta}. 
\eeq
There are $4p$ linearly independent solution, whose asymptotic forms for large $\beta$ are given by
\beq
\frac{1}{\beta} \frac{Z_p}{Z_0} ~\approx~ \beta^q e^{- (2 \epsilon \pm s) m \beta} ~\mbox{or}~ \beta^q e^{\pm s m \beta},  
\\ \notag
(q=0,\cdots,p-1). 
\eeq
Since the leading behavior of the $p$-bion contribution for large $\beta$ should be $Z_p/Z_0 \approx \beta^p$, 
the above asymptotic solutions imply that the term with $s=0$ gives the leading contribution for large $\beta$. 
In the following, we only consider the term with $s=0$. 
For fixed values of $\sigma$, the saddle points of $\mathcal V_i$ and $\tilde{\mathcal V}_i$ are
\beq
z_i = \log \left(\frac{4m}{g^2} \frac{1}{\epsilon_i - i \sigma} \right) + \pi i (2 l_i -1), 
\notag \\
\tilde z_i = \log \left(\frac{4m}{g^2} \frac{1}{\epsilon_i - i \sigma} \right) + \pi i (2 \tilde l_i -1),
\eeq
where $l_i,\,\tilde l_i $ are integers labeling the saddle points. 
It is convenient to shift the integration contour for $\sigma$ 
so that ${\rm Re} (\epsilon_i - i \sigma) > 0$ for all $i$. 
Then, the integration over the thimble $\mathcal J_{l_i, \tilde l_i}$ associated with the saddle point labeled by $(l_i, \tilde l_i)$ gives  
\beq
&&\int_{\mathcal J_{l_i, \tilde l_i}} dz_i \wedge d \tilde z_i \, e^{-\mathcal V_i - \tilde{\mathcal V}_i} =
\notag \\
&& 
\left( \frac{2m}{g^2} e^{ \pi i (l_i+\tilde l_i -1)} \right)^{i \sigma - \epsilon_i} 
\Gamma \left( \frac{\epsilon_i - i\sigma }{2} \right)^2 .
\eeq
The saddle points which contribute to the partition function can be determined by the Lefschetz thimble method. 
In \cite{Fujimori:2016ljw}, we have shown that when $\epsilon_i - i \sigma$ is a positive real number, 
the thimbles which contribute to the partition function are 
\beq
{\mathcal C}_{\mathbb R}=
\left\{ \begin{array}{l}
{\mathcal J}_{1,1}- {\mathcal J}_{1,0} \hs{10} {\rm Im}g^{2}=+0  \\ 
{\mathcal J}_{0,1}- {\mathcal J}_{0,0}  \hs{10} {\rm Im}g^{2}=-0
\end{array} 
\right. .
\eeq
As long as ${\rm Re} \, (\epsilon_i - i \sigma) > 0$, 
we can show that the same thimbles have contributions to the partition function. 
Thus, we obtain
\beq
I_i ~=~ 
\frac{2m}{g^2} \left( \frac{2m}{g^2} e^{\pm \frac{\pi i}{2}} \right)^{i \sigma- \epsilon_i} 
\frac{\displaystyle \Gamma \left( \frac{\epsilon_i - i\sigma }{2} \right)}{\displaystyle \Gamma \left( 1 - \frac{\epsilon_i - i \sigma }{2} \right)},  
\eeq
where we have used the reflection formula for the gamma function
\beq
\sin (\pi x) \, \Gamma(x) = \frac{\pi}{\Gamma(1-x)}.
\eeq
Then, the contour integral for the $p$-bion contribution 
\beq
\frac{Z_p}{Z_0} \approx \frac{m \beta}{p} e^{-\frac{2pm}{g^2}} \int \frac{d\sigma}{2\pi} e^{- i  \sigma m \beta} \prod_{i=1}^{2p} I_i |_{s=0},
\eeq
can be evaluated by picking up the poles at $\sigma = -2i k$ and $\sigma = -2i(\epsilon + k)$~($k\in \Z_{\geq 0}$).
In the $\beta \rightarrow \infty$ limit, 
the $p$-th order pole at $\sigma = 0$ gives the leading order term Eq.\,\eqref{eq:Zp}
\beq
\frac{Z_p}{Z_0} &\approx& 
-\frac{i m \beta}{p} e^{-\frac{2pm}{g^2}} \underset{\sigma=0}{\rm Res} 
\left[ e^{- i m \sigma \beta} \prod_{i=1}^{2p} I_i |_{s=0} \right] \notag \\ 
&=& - \frac{i m \beta}{p!} e^{-\frac{2pm}{g^2}} 
\lim_{\sigma \rightarrow 0} 
\left( \frac{\p}{\p \sigma} \right)^{p-1} 
\notag \\
&&\times \Bigg[ \frac{8i m^2}{g^4} e^{-\frac{i \sigma m \beta}{p}}
\left( \frac{2m}{g^2} e^{\pm \frac{\pi i}{2}} \right)^{2 (i \sigma-\epsilon)} 
\notag \\
&& \times \frac{\Gamma \left( \epsilon - \frac{i\sigma }{2} \right)}{\Gamma \left( 1- \epsilon + \frac{i\sigma }{2} \right)} 
\frac{\Gamma \left( 1- \frac{i\sigma}{2} \right)}{\Gamma \left( 1 + \frac{i\sigma }{2} \right)} \Bigg]^p .
\eeq
The leading order term Eq.~(\ref{eq:p-bion_leading}) is
\beq
\frac{Z_p}{Z_0} &\approx& \frac{1}{p!} \left[  \frac{ 2 m \beta \Gamma \left( \epsilon \right)}{\Gamma \left( 1- \epsilon \right)} e^{-\frac{2m}{g^2} \mp \pi i \epsilon}  \left( \frac{2m}{g^2} \right)^{2 (1-\epsilon)} \right]^p . \notag \\
\eeq 
This is consistent with the dilute gas approximation. 
In the supersymmetric case $\epsilon=1$, $Z_p/Z_0$ vanishes due to the factor 
$1/\Gamma(1-\epsilon)$. 
In the near SUSY case, we obtain
\beq
\lim_{\epsilon \rightarrow 1} \frac{\p}{\p \epsilon} \frac{Z_p}{Z_0} &\approx& 2 m \beta e^{-\frac{2pm}{g^2}},
\eeq
where we have used
\beq
\lim_{x \rightarrow 0} \p_x \frac{1}{\Gamma (x)} = 1. 
\eeq
Then we obtain 
\beq
\lim_{\epsilon \rightarrow 1} \p_\epsilon E &=&
\lim_{\epsilon \rightarrow 1} \p_\epsilon 
\left[ E_{\rm pert} - \lim_{\beta \rightarrow \infty} \frac{1}{\beta} \log \left( 1 + \sum_{p=1}^\infty \frac{Z_p}{Z_0} \right) \right] 
\notag \\
&=& \lim_{\epsilon \rightarrow 1} \p_\epsilon E_{\rm pert} - \sum_{p=1}^\infty e^{-\frac{2pm}{g^2}} \left( 2m + \mathcal O(g^2) \right). \notag \\
\eeq
This is consistent with the exact result.
Using the relation
\beq
&&\frac{1}{p!} \lim_{\epsilon \rightarrow 1} \p_\epsilon^2 \lim_{\sigma \rightarrow 0} \p_\sigma^{p-1} \left[ \frac{X}{\Gamma \left( 1 - \epsilon + \frac{i\sigma }{2} \right)} \right]^p 
\\ \notag 
&&= \lim_{\epsilon \rightarrow 1} \lim_{\sigma \rightarrow 0} 
p \left( \frac{i}{2} X \right)^{p-1} \Big[ (p+1) \gamma - 2 (p-1) i \p_\sigma - 2 \p_\epsilon \Big] X, 
\eeq
we can show that
\beq
&&\frac{1}{2} \lim_{\epsilon \rightarrow 1} \p_\epsilon^2 \frac{Z_p}{Z_0} \approx \\
&& -2 m \beta e^{-\frac{2 p m}{g^2}} \left[ 2 p^2 \left( \gamma + \log \frac{2m}{g^2} \pm \frac{\pi i}{2} \right) - (p-1)m \beta \right]. \notag 
\eeq
Therefore, the second order coefficient of the ground state energy in Eq.~(20) is given by
\beq
&&\frac{1}{2} \lim_{\epsilon \rightarrow 1} \p_\epsilon^2 E = \frac{1}{2} \lim_{\epsilon \rightarrow 1} \lim_{\beta \rightarrow \infty} \left( - \frac{1}{\beta} \frac{Z \p_\epsilon^2 Z - (\p_\epsilon Z)^2}{Z^2} \right)
\notag \\
&&= \sum_p 4 m e^{-\frac{2 p m}{g^2}} p^2 \left[ \gamma + \log \frac{2m}{g^2} \pm \frac{\pi i}{2} + \mathcal O(g^2) \right].
\notag \\
\eeq


\section{Perturbation series on trivial vacuum}
\begin{figure}[t]
\includegraphics[width=70mm]{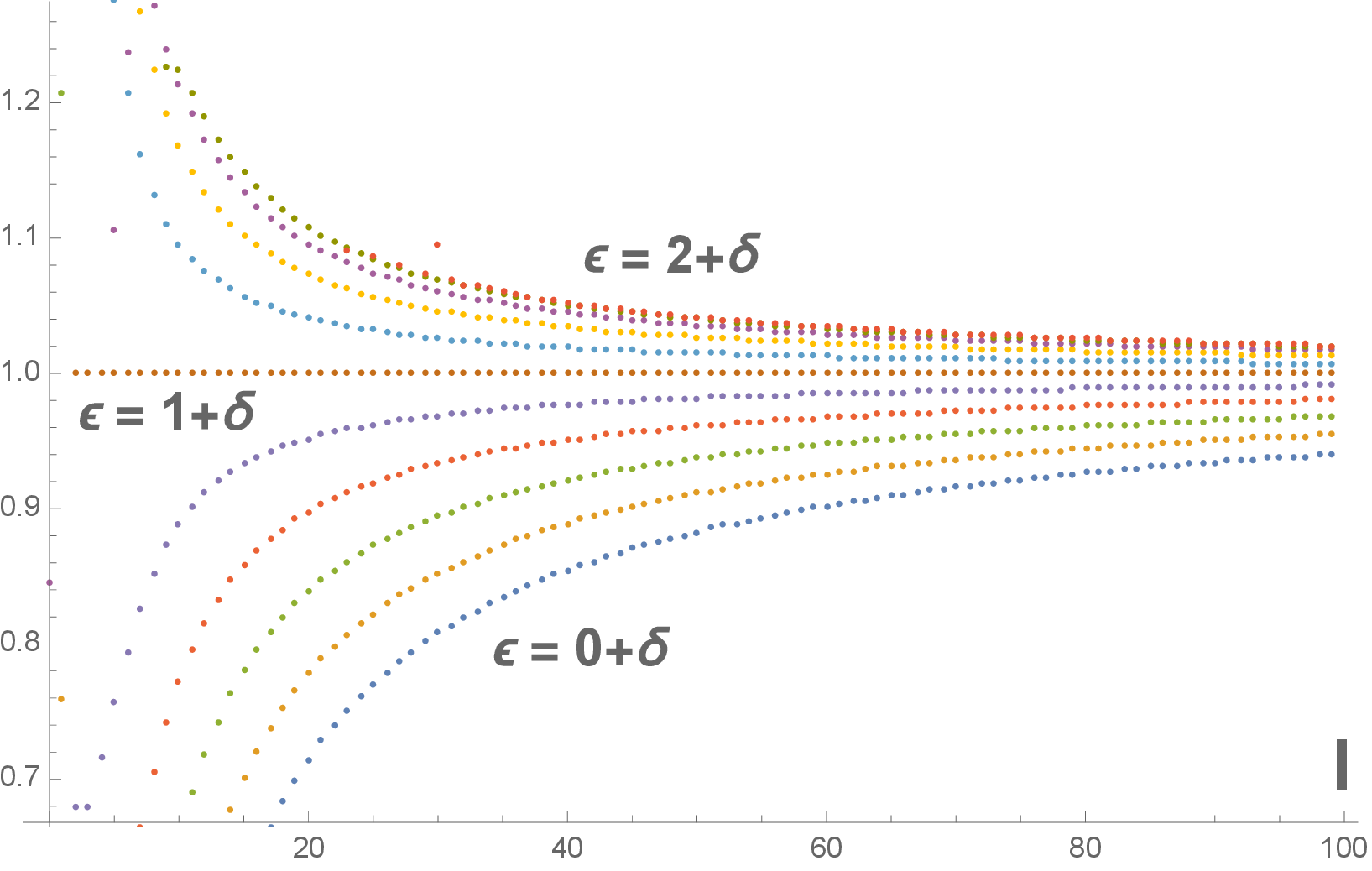}
\caption{The asymptotic behavior of the ratio $A_l/\left[ \frac{1}{2^{l-1}} \frac{\Gamma(l+2(1-\epsilon))}{\Gamma(1-\epsilon)^2}\right]~(l \leq 100)$ for $0 \leq \epsilon \leq 2$ ($\epsilon = n/5,~n=0,\cdots,10$). $\delta$ is a regularization parameter ($\delta = 10^{-10}$).}
\label{fig:asymptotic}
\end{figure}

In this section we derive the perturbative part of the ground state energy by using the Bender-Wu method. 
Since the ground state is invariant under the phase rotation $\varphi \rightarrow e^{i b}\varphi$, 
the corresponding wave function $\Psi$ a function of $|\varphi|$. 
By redefining the wave function and the coordinate as
\beq
\Psi = e^{-x^2} \psi (x), \hs{10} 
|\varphi| = \eta x , \hs{10} 
\eta \equiv \frac{g}{\sqrt{m}}.
\eeq
The Schr\"odinger equation can be rewritten as
\beq
m \Bigg[ - \frac{1}{4} (1+\eta^2 x^2)^2 \left\{ \frac{\p^2}{\p x^2} +(1-4x^2) \frac{1}{x} \frac{\p}{\p x} \right\} 
\notag \\
+ V(x) \Bigg] \psi = E \psi,
\eeq
where the potential is
\beq
V(x) = (1-x^2) (1+ \eta^2 x^2)^2 + \frac{x^2}{(1+\eta^2 x^2)^2} 
-\epsilon \frac{1-\eta^2 x^2}{1+\eta^2 x^2}. \notag \\
\eeq
Let us expand the energy and the wave function with respect to $\eta$
\beq
\frac{E}{m} = \sum_{l=0}^\infty A_l \eta^{2l}, \hs{10} 
\psi = \sum_{l=0}^\infty \psi_{l}(x) \eta^{2l}.
\eeq
Then, the Schr\"odinger equation $(\tilde H - E) \psi = 0$ can be expanded as
\beq
0 &=& \frac{1}{4} \sum_{i=0}^4 \ba{cc} 4 \\ i \ea x^{2i} \Big[ \psi_{l-i}'' + (1-4x^2) \frac{1}{x} \psi_{l-i}'
\notag \\
&-& 4(1-x^2) \psi_{l-i} \Big]  \notag \\
&+& \sum_{i=0}^{l} A_l \left( \psi_{l-i} + 2 x^2 \psi_{l-i-1} + x^4 \psi_{l-i-2} \right) 
\notag \\
&+& (\epsilon-x^2) \psi_l - x^4 \epsilon \psi_{l-2},
\eeq
where $\psi_l=0$ for $l<0$. Setting $\psi_0=1$, we can solve these equations order by order. 
It is not difficult to show that $\psi_l$ are polynomials of the form
\beq
\psi_l = \sum_{k=0}^{2l} B_{l,k} x^{2k}. 
\eeq
We can always fix the normalization of the wave function as $\Psi(x=0) = 1$,
i.e. $B_{0,0}=1$,  $B_{l,0} = 0~(l \not = 0)$.
The Schr\"odinger equation reduces to
\beq
0 &=& \sum_{i=0}^4 \ba{c} 4 \\ i \ea \Big[ (k-i+1)^2 B_{l-i,k-i+1} 
\notag \\
&-& (2k-2i+1) B_{l-i,k-i} + B_{l-i,k-i-1} \Big] \notag \\
&+& \sum_{i=1}^{l} A_i (B_{l-i,k} + 2 B_{l-i-1,k-1} + B_{l-i-2,k-2})
\notag \\
&-& B_{l,k-1} + \epsilon(B_{l,k}-B_{l-2,k-2}), 
\eeq
where $B_{l,k}=0$ if $l<0$, $k<0$, $k>2l$. 
As shown in Fig.\,\ref{fig:asymptotic}, 
the asymptotic behavior Eq.~(\ref{eq:Al}) for $0 < \epsilon < 2$  is consistent with 
\beq
A_l \sim - \frac{1}{2^{l-1}} \frac{\Gamma(l+2(1-\epsilon))}{\Gamma(1-\epsilon)^2}. 
\eeq
The Borel resummation of the right hand side gives
\beq
&&- m \sum_{l=1}^\infty \frac{1}{2^{l-1}} \frac{\Gamma(l+2(1-\epsilon))}{\Gamma(1-\epsilon)^2} \eta^{2l}
\notag \\
 &\overset{\rm Borel}{=}& 
- \frac{g^2}{\Gamma(1-\epsilon)^2} \int_{0}^\infty dt \, e^{-t} \sum_{l=0}^\infty t^{2(1-\epsilon)} \left( \frac{\eta^2}{2} t \right)^l \notag \\
&=& \frac{2 m}{\Gamma(1-\epsilon)^2} \int_{0}^\infty dt \, e^{-t} \frac{t^{2(1-\epsilon)}}{t - \frac{2m}{g^2}}. 
\eeq
Therefore the imaginary ambiguity Eq.~(22) from the perturbative part is
\beq
{\rm Im} \, E_{0} = \mp \frac{2\pi m}{\Gamma(1-\epsilon)^2} \left( \frac{g^2}{2m} \right)^{2(\epsilon-1)} e^{-\frac{2m}{g^2}}.
\eeq


\end{document}